\documentclass[11p]{article}
\usepackage{hyperref}

\RequirePackage{geometry}
\renewcommand{\title}[1]{%
  \par\begingroup
  \centering\LARGE\bfseries #1\par
  \endgroup
  \vspace{1em}%
}

\renewcommand{\author}[1]{%
  \par\begingroup
  \centering\normalsize #1\par
  \endgroup
  \vspace{0.5em}%
}

\providecommand{\affil}[1]{%
  \par\begingroup
  \centering\small #1\par
  \endgroup
}

\providecommand{\email}[1]{%
  \par\begingroup
  \centering\small\texttt{#1}\par
  \endgroup
}

\providecommand{\orcid}[1]{}

\providecommand{\keywords}[1]{%
  \par\medskip
  \noindent\textbf{Keywords:} #1\par
}

\providecommand{\articletype}[1]{}

\providecommand{\ack}[1]{\section*{Acknowledgments}#1}

\usepackage{cite}

\usepackage[T1]{fontenc}
\usepackage{lmodern} 

\newtheorem{theorem}{\bf Theorem}[section]

\usepackage{graphicx}
\usepackage{dcolumn}
\usepackage{bm}
\usepackage{mathtools}
\usepackage{amssymb}
\usepackage{comment}
\usepackage{cleveref}
\usepackage[shortlabels]{enumitem}
\usepackage[title]{appendix}
\usepackage{fix-cm} 

\newcommand{\de}{\,\mathrm{d}}

\newcommand{\R}{\mathbb{R}}

\newcommand{\Z}{\mathbb{Z}}

\newcommand{\abs}[1]{\left|#1\right|}

\newcommand\inner[1]{\left\langle#1\right\rangle}
\newcommand{\innerdots}{\inner{\;\cdot\;,\;\cdot\;}}

\newcommand{\secref}[1]{Sec.~\ref{#1}}
\newcommand{\appref}[1]{App.~\ref{#1}}

\begin{document}

\articletype{Paper} 

\title{When can phonon modes be trapped by arbitrarily weak local defects?  Rigorous sufficient conditions in a monatomic lattice}

\author{Jaden Thomas$^1$\orcid{0000-0000-0000-0000}, Rodrigo Arrieta$^{2,*}$\orcid{0009-0006-4461-8189}, Shu-Ching Yang$^3$, Arthur J.
Parzygnat$^4$\orcid{0000-0002-7264-3991} and Steven G. Johnson$^{1,2}$\orcid{0000-0001-7327-4967}}

\affil{$^1$Department of Physics, Massachusetts Institute of
Technology, Cambridge, Massachusetts 02139, USA}

\affil{$^2$Department of Mathematics, Massachusetts Institute
of Technology, Cambridge, Massachusetts 02139, USA}

\affil{$^3$Mathematical Institute, University of Oxford, Oxford,
UK}

\affil{$^4$Experimental Study Group, Massachusetts Institute of
Technology, Cambridge, Massachusetts 02139, USA}

\affil{$^*$Author to whom any correspondence should be addressed.}

\email{rarrieta@mit.edu}

\keywords{localization, defects, lattice dynamics,
phonons}

\begin{abstract}
Although many studies have shown that localized defects in vibrational lattices can trap localized phonon modes for \emph{particular} geometries, there has been no general proof of localization for \emph{arbitrary} localized defects of arbitrarily weak strength.  This is distinct from the problem of Anderson localization, where random perturbations are spread across the entire lattice and one aims to prove whether \emph{all} states become localized.   Here, we study the former problem for monatomic 1d and 2d crystals with nearest-neighbor harmonic interactions, corresponding in 2d to square lattices of masses and springs, and prove that arbitrary localized perturbations (such as any finite set of defect masses) that decrease the net mass by \emph{any} amount must lead to localized states. In contrast, the analogous statement is generally false in 3d. Our proof employs a straightforward variational method, inspired by analogous results for the Schr{\"o}dinger and Maxwell equations, that should be extensible to other discrete systems.
\end{abstract}

\section{Introduction}
It is well known that localized perturbations (``defects'') in vibrational lattices can trap localized phonon modes, a fact that has important consequences for thermal and electronic transport; this localization phenomenon has been demonstrated in countless semi-analytical and computational studies (in 1d~\cite{Montroll1955, Teramoto1960, movchan2007, Zalipaev2008, Andrianov2021}, 2d~\cite{movchan2007, Osharovich2012, Colquitt2013, Kutsenko2014}, and 3d~\cite{maradudin1963theory, Bttger1983, Stoneham2001, Lipkin2006} lattices), as well as in laboratory experiments~\cite{Barker1975}.  Analogous localization phenomena are also well known to occur in continuous-wave systems, such as quantum bound states in the Schr{\"o}dinger equation~\cite{Landau} or optical trapping/guiding in the Maxwell equations~\cite{joannopoulos2008molding}. However, the Schr{\"o}dinger and Maxwell cases benefit from a theoretical foundation that is lacking in the phonon case: complementing solutions of bound modes in \emph{specific} geometries, there are general proofs that a broad class of \emph{arbitary} perturbations, no matter how weak, must trap localized modes in the 1d and 2d Schr{\"o}dinger~\cite{simon1976bound, picq1982, yang1989simple, Prodan2006, Hundertmark2007, Frank2008, parzygnat2010sufficient} and Maxwell~\cite{Bamberger1990, Urbach1996, lee2008rigorous} cases (whereas in 3d the defect must be sufficiently strong, as can be shown by explicit counter-examples~\cite{Brandt2012}). To our knowledge, no such proof currently exists for discrete vibrational lattices: all previous results analyzed only solutions for particular defects (e.g.~perturbing a given mass) in particular lattices. Our goal in this paper is to begin filling that gap by proving phonon localization for \emph{arbitrary} localized defects (defined by finite $\ell^1$ norm) in prototypical 1d and 2d lattices.

It is important to distinguish the type of localization we study from Anderson localization: the latter involves perturbations that are \emph{not} localized, but are instead distributed randomly throughout the entire lattice, in which case \emph{all} states can become localized under certain conditions~\cite{Stolz2011}. Indeed, Anderson localization of all states by arbitrarily weak lattice-wide disorders has been proved for discrete 2d lattices in certain cases related to our topic, such as near a band edge for a discrete Schr{\"o}dinger equation~\cite{Stolz2011}, as well as in 1d monatomic vibrational lattices~\cite{Matsuda1970}. 

In this paper, we present a general proof of the existence of localized phonon modes produced by any ``light'' (net-mass reduced) finite-norm perturbation of defect masses in 1d (\secref{sec:1d}) and 2d-square (\secref{sec:2d}) discrete monatomic lattices, directly inspired by analogous theorems previously proved in continuous-wave systems~\cite{simon1976bound, picq1982, yang1989simple, Prodan2006, Hundertmark2007, Frank2008, parzygnat2010sufficient,Bamberger1990, Urbach1996, lee2008rigorous}. Our result complements numerous past numerical and semi-analytical studies of discrete-lattice localization by \emph{specific} defect geometries~\cite{Montroll1955, Teramoto1960, movchan2007, Zalipaev2008, Andrianov2021, Osharovich2012, Colquitt2013, Kutsenko2014}, as well as numerical~\cite{Savin2013, Jiang2010, RodriguezNieva2012, Islam2014} and semi-analytical~\cite{Adamyan2010, Savin2017} studies of the important effects of defects on phonon/thermal transport in 2d vibrating lattices (such as graphene). Technically, our proofs employ a variational method, adapting a trial function proposed in the Schr{\"o}dinger case by Yang and de~Llano~\cite{yang1989simple} for the more challenging case of 2d localization, which was subsequently generalized to other wave systems~\cite{lee2008rigorous,parzygnat2010sufficient}. Here, we consider the simplest monatomic lattices with nearest-neighbor harmonic coupling and out-of-plane motion, but we expect that similar theorems will hold in more general 1d and 2d lattices (but \emph{not} in 3d, where only sufficiently strong defects can localize bound states~\cite{maradudin1963theory, Bttger1983, Stoneham2001, Lipkin2006}). Just as the original Schr{\"o}dinger proofs were extended to arbitrary periodic potentials~\cite{Hundertmark2007, Frank2008, parzygnat2010sufficient}, other wave equations~\cite{Bamberger1990, lee2008rigorous}, and localization within band gaps~\cite{Prodan2006, parzygnat2010sufficient}, we believe that our approach should be generalizable to other vibrating lattices (e.g. multi-atom unit cells) and to gaps.  Moreover, our proof relies on simple constructions---with spectral-theory technicalities deferred to the appendices---making it accessible to a broad audience of physicists and attractive to generalize (\secref{sec:conclusion}).

There is a long history of proofs of localization by ``defects'' (\emph{localized} perturbations) in wave systems, especially for the case of Schr{\"o}dinger's equation, with a key question being whether localized solutions (bound states) arise for \emph{arbitrarily weak} defects. For localization by an attractive potential in vacuum for the Schr{\"o}dinger equation, the 1d  proof is at the level of an undergraduate homework problem~\cite{Landau}, but the 2d case was not proved until a landmark 1976 paper by Simon~\cite{simon1976bound}. Much simpler variational proofs in 2d were later discovered~\cite{picq1982, yang1989simple}, and this approach was generalized to prove localization in optical fibers~\cite{Bamberger1990}, photonic-crystal waveguides~\cite{lee2008rigorous}, and periodic Schr{\"o}dinger potentials~\cite{parzygnat2010sufficient}. In all of these cases, the defect pushes an eigenvalue below the minimum of the continuous spectrum (of allowed energies/frequencies in the bulk medium), but in periodic media it is also well known that localized states can occur within band gaps in the interior of the spectrum~\cite{joannopoulos2008molding}. Kuchment and Ong~\cite{Kuchment2003} proved gap localization for sufficiently strong defects, but eventually 1d~\cite{Prodan2006, parzygnat2010sufficient} and 2d~\cite{parzygnat2010sufficient} localization were proved for arbitrarily weak defects in gaps for the periodic Schr{\"o}dinger case. 

These results motivated us to construct a variational proof in discrete phonon lattices, in which spatial differential equations are replaced by difference equations, employing an analogous variational method in 1d and 2d.  In the phonon case, for a monatomic lattice with a single degree of freedom per mass (e.g.~out-of-plane motion in 2d), the continuous spectrum of the bulk medium consists of a single band bounded \emph{above} by a maximum frequency~\cite{Ashcroft}, and our variational proof shows that an arbitrary ``light'' defect (net reduced mass) pushes an eigenvalue beyond the extremum of the continuous spectrum, hence localizing a bound state.  The locality of the defect, and hence preservation of the continuous spectrum (\appref{app:Weyl}), is enforced by requiring the mass changes to have a finite absolute sum \eqref{eq:cond2_1D}. (It is well known that any eigenvalue lying outside the continuous spectrum must correspond to a localized state, because the bulk Green's function is exponentially decaying at such frequencies; this has been shown in general by contour-integration methods in the phonon-lattice case~\cite{Maradudin1965}.)  The underlying physical intuition is that reducing mass causes the vibrational frequency to increase, allowing a defect mode to oscillate faster than the upper frequency cutoff of the bulk medium. We note, however, that our conditions are sufficient for localization, but not necessary: strong perturbations may induce localization even if  the net mass is not decreased. 

\begin{figure}[t]
    \centering
    \includegraphics[width=1.0\linewidth]{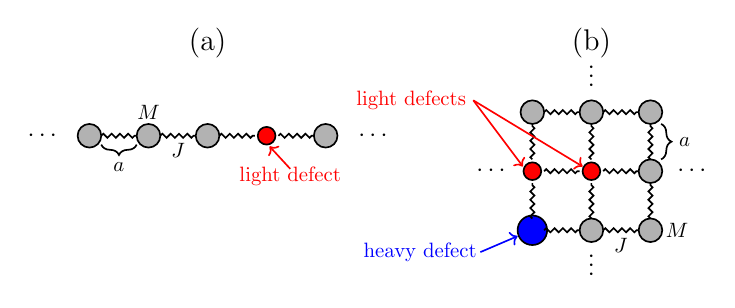}
    \caption{Schematic  monatomic lattices with masses $M$, spring constants $J$ (harmonic nearest-neighbor interactions), and period $a$. Localized perturbations: ``light'' ($< M$, red) and ``heavy'' ($> M$, blue) defect masses. (a)~1d lattice. (b)~2d-square lattice.}
    \label{fig:fig1}
\end{figure}

\section{Proof of 1d localization}
\label{sec:1d}

As in the Schr{\"o}dinger case, our proof of localization in 1d is considerably simpler than in 2d, although both rely on similar variational ideas. It is therefore useful to understand the 1d case before proceeding to 2d in \secref{sec:2d}.  Here, we consider an infinite one-dimensional lattice of masses and springs with nearest-neighbor interactions, as depicted in Fig.~\ref{fig:fig1}(a). We will prove that modifying any number
of masses by a finite total amount [Eq.~{\ref{eq:cond2_1D}], as long as there is a net decrease in their overall sum [Eq.~{\ref{eq:cond1_1D}], leads to the emergence
of at least one localized vibrational mode, such as the examples shown in Fig.~\ref{fig:examples1d}.

\begin{figure}[t]
\centering
\includegraphics[scale=0.35]{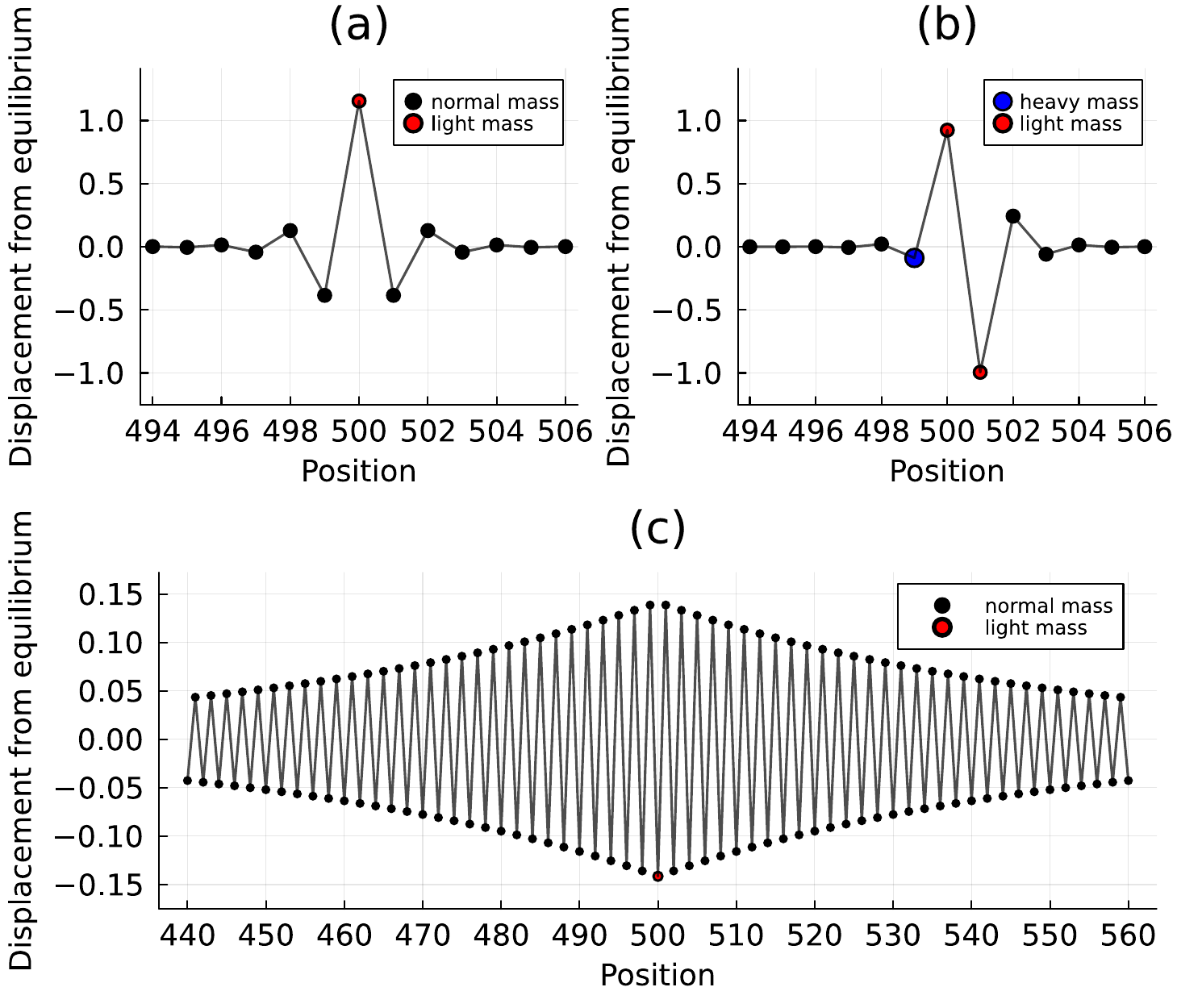}
\caption{Simple 1d examples of perturbations and their corresponding localized modes. (a) Single light-mass perturbation ($M_{\textnormal{light}} = 0.5M$); (b) Two light-mass, one heavy-mass perturbation ($M_{\textnormal{light}} = 0.5M$, $M_{\textnormal{heavy}} = 2M$); (c) single light-mass weak perturbation ($M_{\textnormal{light}} = 0.99M$).  Eigenvectors were computed numerically by a Lanczos method~\cite{Trefethen97,KrylovKit} from a sparse-matrix representation of $\hat{T}$ truncated to a finite supercell of 1000 masses (with Dirichlet boundaries $u_0 = u_{1001}=0$), much larger than the localization length of the bound modes.}
\label{fig:examples1d}
\end{figure}

\subsection{Unperturbed 1d monatomic lattice}
To begin, we review the standard analysis of the spectrum (dispersion relation) of the unperturbed periodic lattice~\cite{Ashcroft}. Let all atoms have mass $M$, separated by the unit-cell period $a$, and let the springs between atoms have elastic constant $J$. We denote the displacement of the $n$-th atom from the equilibrium position by $u_n$. The equation of motion for the $n$-th atom (for either longitudinal or linearized transverse displacements) is:
\begin{align}
M \frac{\partial^{2} u_{n}}{\partial t^{2}}(t)=-J\left[2 u_{n}(t)-u_{n+1}(t)-u_{n-1}(t)\right]\,.
\label{eq:eom}
\end{align}
(The $[\,\cdots\,]$ expression on the right is a discrete Laplacian/graph Laplacian~\cite{Merris1995}, and is also proportional to a finite-difference approximation for $-d^2/dx^2$~\cite[\S25.3.23]{abramowitz+stegun}.)
By periodicity, time-harmonic solutions take the Bloch-wave form
\begin{align}
u_{n}(t)=\tilde{u} e^{i(k a n-\omega t)}\,, \label{eq:normal_mode}
\end{align}
where $k$ is the wave number, $\omega > 0$ is the angular frequency, and $\tilde{u}$ is the wave amplitude. Substituting~(\ref{eq:normal_mode}) into~(\ref{eq:eom}) yields the dispersion relation relating $\omega$ and $k$~\cite{Ashcroft}:
\begin{align}
\omega=\pm \sqrt{\frac{4 J}{M}} \sin (k a / 2)\, .
\label{eq:disp_rel}
\end{align}
A plot of~(\ref{eq:disp_rel}) is shown in
Fig.~\ref{fig:fig2}. Note that the continuous spectrum of bulk-lattice frequencies $\omega \in [-\sqrt{4J/M}, +\sqrt{4J/M}]$ is bounded above (and below). 

\subsection{Perturbed 1d monatomic lattice}

Next, we consider perturbed masses $M_n = M + \Delta M_n$ (leaving the spring constants $J$ unmodified). The equation of motion for the $n$-th atom is then
\begin{align}
M_n \frac{\partial^{2} u_{n}}{\partial t^{2}} (t)=-J\left[2 u_{n}(t)-u_{n+1}(t)-u_{n-1}(t)\right] \, .
\label{eq:eom_pmass}
\end{align}
Again, we seek time-harmonic solutions of the form $u_{n}(t)={u}_{n} e^{-i \omega t}$, where $u_n$ encodes the spatial dependence. Substituting into (4), we obtain
\begin{align}
    \frac{J}{M_n}\left(2 {u}_{n}-{u}_{n-1}-{u}_{n+1}\right)=\omega^{2} {u}_{n} \, .
    \label{eq:eom_pmass_timeharmonic1d}
\end{align}
Equivalently, we can express the solution as an infinite-dimensional complex vector $u \in \mathbb{C}^{\mathbb{Z}}$ (i.e., bi-infinite sequences of complex numbers):
\begin{align}
    {u}=\begin{pmatrix}\cdots & {u}_{-2} & {u}_{-1} & {u}_{0} & {u}_{1} & {u}_{2} & \cdots \end{pmatrix}^{\top} \, ,
\end{align}
where $\top$ denotes transposition. Eq.~(\ref{eq:eom_pmass_timeharmonic1d}) can then be rewritten as the eigenproblem
\begin{align}
    \hat{T} {u}=\omega^{2} {u} \, , \label{eq:pmass_eigenvalue}
\end{align}
where $\hat{T}: \mathbb{C}^{\mathbb{Z}} \rightarrow \mathbb{C}^{\mathbb{Z}}$ is defined by
\begin{align}
    (\hat{T} u)_{n}\coloneqq\frac{J}{M_n}\left(2 u_{n}-u_{n-1}-u_{n+1}\right)\,, \label{eq:t_def}
\end{align}
for $n \in \mathbb{Z}$ and $u \in \mathbb{C}^{\mathbb{Z}}$, which is Hermitian for a weighted inner product (given below). (For infinite lattices where some care is required with the infinite sums, the self-adjointness of $\hat{T}$ is verified in~\appref{app:SAofT}.) 

To prove the existence of a localized solution, it is sufficient to prove that at least one eigenvalue $\omega^2$ of $\hat{T}$ lies outside the continuous spectrum (where the bulk Green's function is exponentially decaying~\cite{Maradudin1965}), as depicted in Fig.~\ref{fig:fig2}.  (Technically, this assumes that the perturbation $\Delta M_n$ is localized enough that it does not change the continuous spectrum $[0, 4J/M]$ of $\hat{T}$; this is proved for our case in~\appref{app:Weyl}.)  Therefore, one must merely bound an eigenvalue $\omega^2 > 4J/M$, which we accomplish below by a variational proof.

\begin{figure}[t]
    \centering    \includegraphics[width=0.8\linewidth]{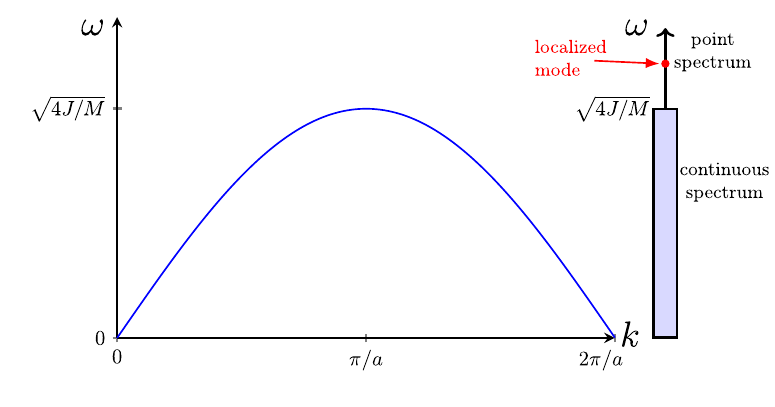}
    \caption{Plot of the well-known dispersion relation~\cite{Ashcroft} for the unperturbed 1d monatomic lattice. When the system is perturbed (e.g.,~decreasing some masses), new discrete frequencies can arise outside of this continuous spectrum, corresponding to localized states.}
    \label{fig:fig2}
\end{figure}

\subsection{Variational proof of 1d localization}
Our proof of the existence of localized vibration modes is not based on an explicit construction of the solution. Instead, it suffices to demonstrate the existence of an eigenvalue $\omega^2$ of $\hat{T}$ that falls outside the continuous spectrum of the bulk dispersion relation, that is $\omega^2 > 4J/M$. 
 
To prove this result, the key tool we employ is the \textit{min--max theorem} (also known as the \textit{variational theorem})~\cite{reed1978iv}: for any bounded self-adjoint operator $\hat{A}$ on a Hilbert space $\mathcal{H}$, its maximum eigenvalue $\lambda_\text{max}$ satisfies $\lambda_\text{max} \geq R_{\hat{A}}\{v\}$ for all $v\in \mathcal{H}$, where $R_{\hat{A}}\{v\} \coloneqq \inner{v, A v}_\mathcal{H} / \inner{v, v}_\mathcal{H}$ is the Rayleigh quotient and $\innerdots_\mathcal{H}$ denotes the inner product on $\mathcal{H}$. 

The operator $\hat{T}$ is indeed bounded and self-adjoint under the weighted $\ell^2$ inner product (see also~\appref{app:SAofT})
\begin{align}
    \inner{u,v}_M \coloneq \inner{u,\hat{M} v} \coloneqq \sum_{n\in\mathbb{Z}} \overline{u}_n M_n v_n \, ,\label{eq:inner_prod_1D}
\end{align}
where $\inner{u,v}$ is the unweighted $\ell^2$ inner product and $\hat{M}$ is the operator that multiplies elementwise by $M_n$.
(Our Hilbert space $\mathcal{H}$ is thus the subspace of $\mathbb{C}^\mathbb{Z}$ with finite $\ell^2$ norm.)
Hence, our problem is reduced to finding an appropriate trial function $v^*$ such that $R_{\hat{T}}\{v^*\} > 4J/M$. The min--max theorem (with the $\innerdots_M$ inner product) then guarantees the existence of an eigenvalue $\omega^2 > 4J/M$ of $\hat{T}$, which in turn implies the existence of a localized solution.

For a 1d perturbed monatomic lattice, we allow the perturbed masses $M_{n} = M + \Delta M_n$ to vary, either being heavier or lighter than the original mass $M$; however, we impose the following three conditions on the perturbation $\{\Delta M_n\}_{n\in\mathbb{Z}}$:
\begin{enumerate}
    \item The masses are nonnegative: $M_n = M + \Delta M_n > 0$, to ensure that~(\ref{eq:inner_prod_1D}) is an inner product.

    \item The net mass is decreased. 
    \begin{align}
        \sum_{n\in\Z} \Delta M_{n}<0\,. \label{eq:cond1_1D}
    \end{align}
    \item The total mass added or removed (the $\ell^1$ norm of $\Delta M$) must be finite.
    \begin{align}
        \sum_{n\in\Z}\left|\Delta M_{n}\right|<\infty \,. \label{eq:cond2_1D}
    \end{align}
\end{enumerate}
(The third condition ensures that the perturbation is localized enough to not affect the ``essential spectrum'' obtained from the Bloch modes, as shown in~\appref{app:Weyl}; it is also used in analyzing limits below.)
Under these conditions, we state the following theorem on the existence of localized vibrational modes in $1$d:
\begin{theorem}
    If the mass perturbations $\{\Delta M_n\}_{n\in\mathbb{Z}}$ satisfy the conditions above, then there exists at least one localized vibrational mode.
\end{theorem}
\textit{Proof}. Since $\hat{T}$ is self-adjoint under the inner product defined in~(\ref{eq:inner_prod_1D}), the min--max theorem guarantees the existence of a localized mode if we show 
\begin{align}
    R_{\hat{T}}\{v^*\} > 4J/M\,,
\end{align}
or equivalently, 
\begin{align}
    \inner{v^*,\hat{T}v^*}_M > \omega^2_\text{max} \inner{v^*,v^*}_M\,, \label{eq:thm_cond}
\end{align} 
for an appropriate trial function with finite norm ($v^* \in \ell^2$), where $\omega^2_\text{max} \coloneqq 4J/M$.

Consider $v^*_n = \alpha^{\abs{n}}$, with $\alpha\in\R$ and $\abs{\alpha} < 1$. The left-hand side of the inequality~(\ref{eq:thm_cond}) is then
\begin{align}
\begin{split}
    \inner{v^*,\hat{T}v^*}_M
    &= \sum_{n\in\Z} \overline{v^*_n} (Tv^*)_n M_n\\
    &= J\sum_{n\in\Z} \overline{v^*_n} \left(2 v^*_{n}-v^*_{n-1}-v^*_{n+1}\right) \\
    &= 2J \frac{1-\alpha}{1+\alpha}\,.
\end{split}
\end{align}
Conversely, the right-hand side is given by
\begin{align}
\begin{split}
    \omega^2_\text{max}\inner{v^*,v^*}_M
    &= \frac{4J}{M}\sum_{n\in\Z} \abs{v^*_n}^2 M_n\\
    &= \frac{4J}{M}\sum_{n\in\Z} \abs{v^*_n}^2 \left[M + \Delta M_n\right]\\
    &= 4J\frac{1+\alpha^2}{1-\alpha^2} + \frac{4J}{M}\sum_{n\in\Z} \alpha^{2\abs{n}} \Delta M_n\,.
\end{split}
\end{align}
Subtracting the first term on the right-hand side from both sides and dividing by $J$ yields the condition:
\begin{align}
    2 \frac{\alpha + 1}{\alpha - 1} >  \frac{4}{M}\sum_{n\in\Z} \alpha^{2\abs{n}} \Delta M_n \, . \label{eq:intermediate_1d}
\end{align}
To prove that this condition holds for some $|\alpha| < 1$, it suffices to show that it is true in the limit $\alpha \to -1^+$ (in which case it must also be true for some $\alpha$ sufficiently close to $-1$). 

The left-hand side of~(\ref{eq:intermediate_1d}) approaches zero as $\alpha \rightarrow -1^+$. The right-hand side of~(\ref{eq:intermediate_1d}), on the other hand, approaches a negative number:
\begin{equation}
\begin{split}
\lim_{\alpha\rightarrow -1^+}  \left(  \frac{4}{M}\sum_{n\in\Z} \alpha^{2\abs{n}} \Delta M_n \right)
    &=  \frac{4}{M}\sum_{n\in\Z} \Delta M_n \left( \lim_{\alpha\rightarrow -1^+} \alpha^{2\abs{n}} \right) \\
    &=  \frac{4}{M}\sum_{n\in\Z} \Delta M_n < 0\,,
\end{split}
\end{equation}
where in the first equality we employed Tannery's theorem~\cite{ismail2005theory} to interchange the limit and the sum, using $\{\Delta M_n\}_{n\in\Z}$ as the dominating sequence and the fact that $\sum_{n\in\Z}\abs{\Delta M_n} < \infty$ by~(\ref{eq:cond2_1D}), while in the last line we used~(\ref{eq:cond1_1D}). Therefore, there exists an $\alpha$ close to~$-1$ for which~(\ref{eq:thm_cond}) holds, completing the proof for $1$d localization.

The motivation for taking this limit is the fact that $\alpha = -1$ gives the bulk Bloch mode~(\ref{eq:normal_mode}) at $k = \pi/a$, corresponding to the edge of the Brillouin zone and the extremum of the spectrum. One expects weakly localized defect modes to resemble this band-edge solution, and a similar ansatz has been employed in other variational proofs~\cite{lee2008rigorous,parzygnat2010sufficient}. We use the same idea in the 2d proof, below.

\section{Proof of 2d localization}
\label{sec:2d}
We now consider the case of an infinite vibrating two-dimensional lattice of masses and springs with out-of-plane motion (in the harmonic/linearized force approximation), as depicted in Fig.~\ref{fig:fig1}(b). We will again prove that modifying any number
of masses by a finite total amount, as long as there is a net
decrease in their overall sum, leads to the emergence
of at least one localized vibrational mode. 

\subsection{Unperturbed 2d lattice, out-of-plane motion}
We begin by reviewing the analysis of the unperturbed lattice~\cite{slepyan2012models,Ashcroft}. Let all atoms have mass $M$, arranged in a square lattice with lattice constant $a$, and let the springs between the masses have elastic constant $J$ for out-of-plane motion. The linearized (small-displacement) equation of motion for the out-of-plane displacement $u_{n,m}$ is then~\cite{slepyan2012models}:
\begin{multline}
    M \frac{\partial^{2} u_{n,m}}{\partial t^{2}}(t) = -J\big[4u_{n,m}(t) - u_{n+1,m}(t) - u_{n-1,m}(t)- u_{n,m+1}(t) - u_{n,m-1}(t)\big] \, , \label{eq:eom2D}
\end{multline}
where $n$ and $m$ index the in-plane unit cells. (As in 1d, the $[\,\cdots\,]$ on the right-hand side is a graph Laplacian of the 2d lattice~\cite{Merris1995}, and is also proportional to a 5-point finite-difference approximation for $-\nabla^2$~\cite[\S25.3.30]{abramowitz+stegun}.) By periodicity, time-harmonic solutions take the Bloch-wave form~\cite{Ashcroft}
\begin{align}
u_{n,m}(t)=\tilde{u} e^{i(k_x a n + k_y a m -\omega t)}\,, \label{eq:normal_mode2D}
\end{align}
where $k_x$ and $k_y$ are the components of the wavevector. Substituting this into~(\ref{eq:eom2D}) yields the dispersion relation~\cite{Ashcroft}:
\begin{align}
\omega = \pm \sqrt{\frac{4 J}{M}\left[\sin^2\left(\frac{k_x a}{2}\right) + \sin^2\left(\frac{k_y a}{2}\right)\right]} \, .
\label{eq:disp_rel2D}
\end{align}
From the dispersion relation, it is clear that the continuous spectrum of $\omega^2$ is bounded above by $\omega_{\textnormal{max}}^2 \coloneqq 8J/M$.

\subsection{Perturbed 2D lattice}

As in 1d, we will now consider perturbed masses, replacing $M$ in~(\ref{eq:eom2D}) above with $M_{n,m} = M + \Delta M_{n,m} > 0$ at each lattice (where $\Delta M_{n,m}$ is sufficiently localized as described in the next section).   Again, this yields an eigenproblem $\hat{T}u=\omega^2 u$ for the time-harmonic solutions $u_{n,m}(t)={u}_{n,m} e^{-i \omega t}$, where the operator $\hat{T}$ is defined by
\begin{equation}
    (\hat{T} u)_{n,m}\coloneqq\frac{J}{M_{n,m}}(4 {u}_{n,m}-{u}_{n+1,m} -{u}_{n-1,m}-{u}_{n,m+1}-{u}_{n,m-1}) \, ,
\label{eq:T_2d}
\end{equation}
for $n,m \in \mathbb{Z}$ and $u \in \mathbb{C}^{\mathbb{Z}^2}$, which is bounded and self-adjoint for an $M_{n,m}$-weighted inner product similar to~(\ref{eq:inner_prod_1D}), as discussed at the end of~\appref{app:SAofT}.  To establish localization, it is again sufficient to demonstrate the existence of an eigenvalue $\omega^2$ of $\hat{T}$ which lies outside of the continuous spectrum, $\omega^2 > 8J/M$. Just as in the 1d case, this can be accomplished by a variational proof; however, the simple exponential trial function employed in the 1d case does not work in 2d, demanding a more complicated trial function. 

\subsection{Variational proof of 2d localization}
Similar to the three conditions in the 1d case from \secref{sec:1d}, we require non-negative masses $M_{n,m} = M +\Delta M_{n,m}  > 0$, a net mass decrease $\sum_{n,m} \Delta M_{n,m} < 0$, and a finite total perturbation $\sum_{n,m} |\Delta M_{n,m}| < \infty$, with the only difference being that the sums are now over two indices $(n,m) \in \mathbb{Z}^2$.  
We state the following theorem on the existence of localized vibrational modes in $2$d:
\begin{theorem}
    If the mass perturbations $\{\Delta M_{n,m}\}_{(n,m)\in\mathbb Z^2}$ satisfy the conditions above, then there exists at least one localized vibrational mode.
\end{theorem}

\noindent
\textit{Proof}. The operator $\hat{T}$ is self-adjoint under the weighted $\ell^2$ inner product
\begin{align}
    \inner{u,v}_M \coloneqq \inner{u,\hat{M}v} \coloneqq  \sum_{n,m\in\mathbb{Z}} \overline{u}_{n,m} M_{n,m} v_{n,m} \, ,\label{eq:inner_prod}
\end{align}
where $\hat{M}$ is again elementwise multiplication by $M_{n,m}$ and $\innerdots$ is the unweighted $\ell^2$ inner product.
As in the 1d case, our problem is reduced to finding an appropriate trial function $v^*$ such that $R_{\hat{T}}\{v^*\} > 8J/M$ in the $\innerdots_M$ inner product. The min--max theorem then guarantees the existence of an eigenvalue $\omega^2 > 8J/M$ outside the continuous spectrum, which in turn implies localization. (As in~1d, \appref{app:Weyl} shows that this continuous spectrum is not changed by the perturbation.)  Explicitly, we must show
\begin{align}
    \inner{v^*,\hat{T}v^*}_M > \omega^2_\text{max} \inner{v^*,v^*}_M \label{eq:thm_cond2D}
\end{align}
for some $v^*$.
Observe that the second-difference operator in $\hat{T}$ can be factorized into a composition of first differences (just as a continuous Laplacian is the divergence of a gradient):
\begin{align}
\label{eq:difference_factorization}
    \hat{T} = \hat{M}^{-1}J \hat{D}^{\dagger}\hat{D} = \hat{M}^{-1}J  \left( \hat{D}_x^{\dagger}\hat{D}_x + \hat{D}_y^{\dagger}\hat{D}_y \right)\,,
\end{align}
where $\hat{D} \coloneqq \begin{pmatrix}\hat{D}_x \\ \hat{D}_y\end{pmatrix}$ maps $\mathbb{C}^{\mathbb{Z}^2}$ to discrete ``gradients'' in $\mathbb{C}^{\mathbb{Z}^2} \oplus \mathbb{C}^{\mathbb{Z}^2}$, with
\begin{equation}
    (\hat{D}_x u)_{n,m} \coloneqq u_{n+1,m} - u_{n,m} 
\end{equation}
and 
\begin{equation}
    (\hat{D}_y u)_{n,m} \coloneqq u_{n,m+1} - u_{n,m}\,.
\end{equation}
Here, $\hat{D}_{x}^{\dag}$ and $\hat{D}_{y}^{\dag}$ are the adjoints with respect to the unweighted $\ell^2$ inner product $\innerdots$ and $\hat{D}^{\dag}=\begin{pmatrix}\hat{D}_{x}^{\dag}&\hat{D}_{y}^{\dag}\end{pmatrix}$. Identity~\eqref{eq:difference_factorization} corresponds to the well-known factorization of a graph Laplacian into a product $D^{\top} D$ via the incidence matrix $D^{\top}$~\cite{Merris1995}. We can then rewrite~(\ref{eq:thm_cond2D}), divided by $J$, as
\begin{equation}
\begin{split}
   \inner{\hat{D}v^*,\hat{D}v^*} &= \inner{\hat{D}_xv^*,\hat{D}_xv^*} + \inner{\hat{D}_yv^*,\hat{D}_yv^*}> \frac{\omega^2_\text{max}}{J} \inner{v^*,v^*}_M\,. \label{eq:cond2D}
\end{split}
\end{equation}
This factorization is also derived in~\appref{app:SAofT}.

Motivated by the analogous trial function of Yang and de~Llano~\cite{yang1989simple}, generalized in several subsequent works~\cite{lee2008rigorous,parzygnat2010sufficient}, we consider the trial function $v^*$ given by 
\begin{align}
    v^*_{n,m} = (-1)^{n + m} \underbrace{e^{-(n^2+m^2+1)^\alpha}}_{f(n,m)}\,,  
\end{align}
where we defined the function $f(n,m) = e^{-(n^2+m^2+1)^\alpha}$ for any $\alpha > 0$, which has the symmetries:
\begin{equation}
    f(n,m) = f(\pm n, \pm m) = f(\pm m, \pm n)  \, .\label{eq:fsymmetry}
\end{equation}
The oscillatory factor $(-1)^{{n}+{m}}$ is included because it corresponds to the ``band-edge'' Bloch solution of the unperturbed $\hat{T}$ that achieves the maximum eigenvalue $\omega_\text{max}^2$. The left-hand side of~(\ref{eq:cond2D}) is 
\begin{align}
    \inner{\hat{D}v^*,\hat{D}v^*} = \sum_{n,m\in\mathbb{Z}} \left( [f(n+1,m)+f(n,m)]^2 + 
    [f(n,m+1)+f(n,m)]^2 \right) \,.
\end{align}
The right-hand side of~(\ref{eq:cond2D}) is 
\begin{align}
    \frac{\omega^2_{\textnormal{max}}}{J}\inner{v^*,v^*}_M = 8\sum_{n,m\in\mathbb{Z}} f(n,m)^2 + \frac{8}{M}\sum_{n,m\in\mathbb{Z}} f(n,m)^2\Delta M_{n,m}\,.
\end{align}
Rearranging~(\ref{eq:cond2D}),  using symmetry~\eqref{eq:fsymmetry} and moving the $8\sum f^2$ term to the left-hand side, we obtain
\begin{align}
    2\sum_{n,m\in\mathbb{Z}} \left( \left[f(n+1,m)+f(n,m)\right]^2 - 4f(n,m)^2 \right) >   \frac{8}{M}\sum_{n,m\in\mathbb{Z}} f(n,m)^2\Delta M_{n,m}\,. \label{eq:cond2}
\end{align}
To show that this inequality holds for some $\alpha > 0$, we prove it in the limit $\alpha \rightarrow 0^+$. First, note that the right-hand side of~(\ref{eq:cond2}) is negative in this limit:
\begin{align}
\lim_{\alpha\rightarrow 0^+}\frac{8}{M}\sum_{n,m\in\Z}f(n,m)^2 \Delta M_{n,m} = \frac{8e^{-2}}{M}\sum_{n,m\in\Z}\Delta M_{n,m} < 0 \,,
\end{align}
where in the second step we used Tannery's theorem~\cite{ismail2005theory} to interchange the limit and the sum, and $\sum\Delta M_{n,m} < 0$ by assumption. Hence, our problem can be reduced to showing that the left-hand side of~(\ref{eq:cond2}) converges to zero as $\alpha\rightarrow 0^+$.

We now define a quantity $S$ equal to the left-hand side of~(\ref{eq:cond2}) and perform some simplifications,
\begin{align}
    S &\coloneqq 2\sum_{n,m\in\mathbb{Z}} \left(\left[f(n+1,m)+f(n,m)\right]^2 - 4f(n,m)^2\right) \nonumber \\
    &= -2\sum_{n,m\in\mathbb{Z}} \left[f(n+1,m)-f(n,m)\right]^2 \leq 0 \,,
\end{align}
where $\sum f(n,m)^2 = \sum f(n+1,m)^2$ was used on the second line to transform $-2f(n,m)^2$ into $-2f(n+1,m)^2$ from the $-4f(n,m)^2$ term.  This transformation relies on the square-summability of $f$, which allows us to individually re-arrange the terms in the sum.
We now show that $S\rightarrow0$ as $\alpha\rightarrow0^+$. 
Using the symmetries~\eqref{eq:fsymmetry} of $f(n,m)$, we fold the 2d sum into the first quadrant, resulting in:
\begin{align}
    S = -8\sum_{n,m\geq 0}\left[f(n+1,m)-f(n,m)\right]^2+4\sum_{n\geq0,}\left[f(n+1,0)-f(n,0)\right]^2 \,.
\end{align}
Using the triangle inequality and bounding the 1d sum by the 2d sum, we have
\begin{equation}
    \abs{S} \leq 12\sum_{n,m\geq 0}\left[f(n+1,m)-f(n,m)\right]^2 \, .
\end{equation}
Hence, it is sufficient to show that this 2d sum vanishes as $\alpha\rightarrow0^+$. To this end, we will bound sums by integrals and then show that those integrals vanish.

We now employ the mean-value theorem to rewrite
\begin{align}
    f(n+1,m)-f(n,m) = \partial_nf(\xi_{n,m},m) \, ,
\end{align}
where $\xi_{n,m}\in (n,n+1)$~\cite{rudin1976principles}. Therefore, 
\begin{align}\label{eq:S0def}
    \frac{\abs{S}}{12}\leq\sum_{n,m\geq 0}\abs{\partial_nf(\xi_{n,m},m)}^2 \, .
\end{align}
Explicitly evaluating the summand for our particular trial function $f$, and denoting $\xi = \xi_{n,m}$, we can simplify:
\begin{equation}
\begin{split}
\abs{\partial_nf(\xi,m)}^2 &= \frac{4\alpha^2 \xi^2}{\left(\xi^2+m^2+1\right)^{2-2\alpha}}\, f(\xi,m)^2\\
&\leq\frac{4\alpha^2 \left(\xi^2+m^2+1\right)}{\left(\xi^2+m^2+1\right)^{2-2\alpha}}\, f(\xi,m)^2\\
&=\frac{4\alpha^2 }{\left(\xi^2+m^2+1\right)^{1-2\alpha}}\, f(\xi,m)^2\\
&\leq\frac{4\alpha^2 }{\left(n^2+m^2+1\right)^{1-2\alpha}}\, f(n,m)^2 \, ,
\end{split}
\label{eq:f_deriv_bound}
\end{equation}
where in the last line we used the fact that the right-hand side is monotonically decreasing in $\xi > n$ when, e.g., $\alpha\in(0,0.25)$. Substituting \eqref{eq:f_deriv_bound} into \eqref{eq:S0def} and using the integral bounds \eqref{eq:2d_bounds_2} (in~\appref{app:integral-bounds}) for monotonically decreasing summands yields:
\begin{align}
\begin{split}
    \frac{\abs{S}}{12}&\leq \sum_{n,m\geq 0} \frac{4\alpha^2 \,f(n,m)^2}{\left(n^2+m^2+1\right)^{1-2\alpha}}\\
    &\leq 4\alpha^2f(0,0)^2 + 2\int_0^\infty \frac{4\alpha^2 \,f(r,0)^2}{\left(r^2+1\right)^{1-2\alpha}} \de r 
    + \int_0^\infty \int_0^\infty \frac{4\alpha^2 \,f(x,y)^2}{\left(x^2+y^2+1\right)^{1-2\alpha}} \de x \de y \\
    &= 4\alpha^2 e^{-2} + 2\int_0^\infty \frac{4\alpha^2 \,e^{-2\left(r^2+1\right)^\alpha}}{\left(r^2+1\right)^{1-2\alpha}} \de r 
    + \frac{\pi}{2} \int_0^\infty \frac{4\alpha^2 \,e^{-2\left(r^2+1\right)^\alpha}}{\left(r^2+1\right)^{1-2\alpha}}\,r \de r \, , \label{eq:S0int1}
\end{split}
\end{align}
where in the last line we rewrote the 2d integral in polar coordinates. The two integrands are almost the same, but the latter has an $r$ factor; to combine the integrals we can employ the following inequality for positive functions $h(r)$, so that we only have a single infinite integral to analyze:
\begin{align}
\begin{split}
    2\int_0^\infty h(r)\de r + \frac{\pi}{2}\int_0^\infty h(r)\,r\de r \leq \left(2+\frac{\pi}{2}\right) \int_0^1 h(r)\de r + \left(2+\frac{\pi}{2}\right) \int_1^\infty h(r)\,r\de r \, .
\end{split}
\end{align}
Applying this inequality to \eqref{eq:S0int1}, we obtain:
\begin{align}
    \frac{\abs{S}}{12}\leq\; 4\alpha^2 e^{-2} + 4\alpha^2\left(2+\frac{\pi}{2}\right)\int_0^1 \frac{ \,e^{-2\left(r^2+1\right)^\alpha}}{\left(r^2+1\right)^{1-2\alpha}} \de r + 4\alpha^2\left(2+\frac{\pi}{2}\right)\int_1^\infty \frac{ \,e^{-2\left(r^2+1\right)^\alpha}}{\left(r^2+1\right)^{1-2\alpha}}\,r \de r \, . \label{eq:S0bound}
\end{align}
The $\int_0^1$ term can be easily shown to vanish as $\alpha\rightarrow 0^+$ by employing the dominated convergence theorem~\cite{rudin1987real} to interchange the limit and the integral, since the integrand is bounded in the $r \in [0,1]$ interval by its $\alpha$-independent value at $r=0$. The $\int_1^{\infty}$ term can be directly evaluated in closed form:
\begin{align}
    4\alpha^2\left(2+\frac{\pi}{2}\right)\int_1^\infty \frac{ \,e^{-2\left(r^2+1\right)^\alpha}}{\left(r^2+1\right)^{1-2\alpha}}\,r \de r \nonumber&= 4\alpha^2\left(2+\frac{\pi}{2}\right)\left[\frac{-1}{8\alpha}\left(2\left(r^2+1\right)^\alpha +1\right)e^{-2\left(r^2+1\right)^\alpha}\right]\Bigg\lvert_{r=1}^{r=\infty} \nonumber \\
    &= \frac{\alpha}{2}\left(2+\frac{\pi}{2}\right)\left(2^{\alpha+1}+1\right)e^{-2^{\alpha+1}} \, ,
\end{align}
which vanishes as $\alpha\rightarrow 0^+$. Therefore, $S$ also vanishes in this limit, completing the proof.

\section{Conclusion and Future Work}\label{sec:conclusion}
This paper presents a rigorous proof establishing sufficient conditions for the existence of localized vibrational modes
in $1$d and $2$d-square monatomic lattices, subject to arbitrary localized perturbations with a net decrease in mass. We believe these results represent an important starting point that can lead to many analogous results for discrete systems.

Although we considered only perturbations in the masses $M$, it should be straightforward to prove an extension to localized perturbations in the spring constants $J$, or in both the spring constants and the masses.  (Lattice-dislocation defects, in which the position of one or more atoms is perturbed~\cite{Hirth1983}, could be expressed in terms of such a change in $J$.) We expect localization to arise from net \emph{increases} in the spring constants ($\sum \Delta J > 0$), since that tends to increase frequency, and more generally the criterion is probably a linear combination of $\sum \Delta J$ and $\sum \Delta M$.   Allowing in-plane ($xy$) motion as well as out-of-plane ($z$) motion introduces more degrees of freedom that may lead to additional localized modes, but the out-of-plane localized states proved in this paper still persist because the (linearized) in-plane and out-of-plane motions are decoupled by the $z=0$ mirror-symmetry plane.  As in the Schr{\"o}dinger~\cite{parzygnat2010sufficient} and Maxwell~\cite{Bamberger1990,lee2008rigorous} cases, the crucial factor is the dimensionality of the \emph{localization}, not of the system, and so we also expect similar localization results to hold in 3d lattices for 1d and 2d localization by plane (2d-periodic) and line (1d-periodic) defects, respectively.  Even more generally, one could consider arbitrary periodic lattices with multiple masses per unit cell, multiple degrees of freedom per mass (motion in several directions), and multiple linear interactions (``springs'').  Such a generalization poses several challenges.  First, there may not be a closed-form expression for the spectrum or band-edge state of the unperturbed lattice, but similar to previous work one could express the theorem in terms of this unknown band-edge state~\cite{lee2008rigorous,parzygnat2010sufficient}.  Second, once there are multiple degrees of freedom per unit cell there will be multiple ``bands'' in the dispersion relation and the possibility of band gaps in the interior of the spectrum~\cite{Ashcroft}---this introduces additional possibilities for localized modes in gaps, which may be possible to study (analogous to previous work on the Schr{\"odinger} case) using a shifted-and-squared operator~\cite{parzygnat2010sufficient}.  Third, to study more general lattices and interactions, it would be desirable to develop a more abstract algebraic framework so that one does not need to laboriously bound every individual term in the Rayleigh quotient.  To address these challenges, we have results under preparation that take a more abstract, graph-theoretic approach, and which prove the existence of an appropriate trial function for each defect rather than employing a generic double-exponential trial function for all defects~\cite{JorgensenAlonJohnson2026-talk}.

Finally, we note that there are other discrete-space wave systems that could benefit from similar analyses of localized defects.  One example is the discrete Schr{\"o}dinger equation on graphs, which have been widely studied for disorder and/or nonlinear effects~\cite{Stolz2011,Damanik2016,Kevrekidis2009}.  Localization in periodic Schr{\"o}dinger graphs should be somewhat easier to study than the phonon case, because for the Schr{\"o}dinger operator $-\nabla^2 + V$ (where $\nabla^2$ is a discrete/graph Laplacian) the potential perturbation $V$ is additive rather than multiplicative with the Laplacian.

\ack{We are grateful to Uma Kausik for numerical simulations that helped to inspire this work, to Carsten Trunk for helpful discussions about operator pencils, and to Mehran Kardar for feedback.  This work was supported in part by the U.S. Army Research Office through the Institute for Soldier Nanotechnologies (Award No.~W911NF-23-2-0121), by the Simons Foundation through the Simons Collaboration on Extreme Wave Phenomena Based on Symmetries, and by the MIT Undergraduate Research Opportunities Program.}

\begin{appendices}

\section{$\hat{T}$ is bounded and self-adjoint}
\label{app:SAofT}

In this Appendix, we assume familiarity with some notions of operator theory and absolute convergence~\cite{Folland99,rudin1976principles}.  We begin with the 1d case; as noted at the end, the analysis of the 2d case is almost the same. 
Fix numbers $M,J>0$ and let $(\Delta M_n)\in\mathbb{C}^{\mathbb{Z}}$ be a bi-infinite sequence satisfying $M + \Delta M_{n} > 0$ for all $n\in\mathbb{Z}$. We let $M_{n}=M+\Delta M_{n}$ and define the operator $\hat{T}: \mathbb{C}^\mathbb{Z}\to \mathbb{C}^\mathbb{Z}$ as in \eqref{eq:t_def}, and similarly $\hat{M}: \mathbb{C}^\mathbb{Z}\to \mathbb{C}^\mathbb{Z}$ is the operator given by $(\hat{M}u)_n = M_n u_n$. Let $\inner{u,v}_M:=\inner{u,\hat{M}v}$ denote the $\hat{M}$-weighted sesquilinear form on $\mathbb{C}^{\mathbb{Z}}$ as in~\eqref{eq:inner_prod_1D}, where $\inner{u,v}: = \sum_n \overline{u_n} v_n$ is the standard (unweighted) $\ell^2$ inner product. Let $\mathcal{H}$ be the Hilbert space consisting of elements $u\in\mathbb{C}^{\mathbb{Z}}$ such that $\lVert u\rVert_M<\infty$, where $\lVert\;\cdot\;\rVert_M$ is the norm induced by this inner product $\innerdots_M$. In this Appendix, we will prove that if $\lim_{n\to\pm\infty}\Delta M_{n}=0$, which is implied by condition~\eqref{eq:cond2_1D}, then $\hat{T}$ defines a bounded self-adjoint operator on $\mathcal{H}$. 

We first prove that $\hat{T}$ is a bounded operator on $\mathcal{H}$, which means that $\lVert\hat{T}\rVert_{M}<\infty$, where the operator norm of $\hat{T}$ is defined as 
\begin{equation}
\lVert \hat{T}\rVert_M=\inf\big\{c\ge0\,:\,\lVert \hat{T}u\rVert_M\le c\lVert u\rVert_M \;\forall u\in\mathcal{H}\big\}.
\end{equation} 
Let $u\in\mathcal{H}$ so that $\lVert u\rVert_M^2=\sum_{n}M_n|u_n|^2<\infty$. Then, by definition of $\hat{T}$, 
\begin{equation}
\lVert \hat{T}u\rVert_M^2=\sum_{n}\frac{J^2}{M_n}|2u_n-u_{n+1}-u_{n-1}|^2.
\end{equation}
Next, let $M_{\text{min}}\coloneqq\inf_{n}M_{n}$ and $M_{\text{max}}\coloneqq\sup_{n}M_{n}$, which are guaranteed to exist and satisfy $M_{\text{max}} \ge M_n \ge M_{\text{min}}>0$ due to the assumptions $M + \Delta M_{n}>0$ and $\lim_{n\to\pm\infty}\Delta M_{n}=0$. Hence, 
\begin{equation}
\label{eq:TuLE2n1n1}
\lVert \hat{T}u\rVert_M^2\le \frac{J^2}{M_{\min}}\sum_{n}|2u_n-u_{n+1}-u_{n-1}|^2.
\end{equation}
To proceed, we expand the absolute value term as 
\begin{equation}
\label{eq:splitLaplacianterms}
\begin{split}
|2u_n-u_{n-1}-u_{n+1}|^2&=4|u_n|^2+|u_{n-1}|^2+|u_{n+1}|^2-2\overline{u}_{n}u_{n-1}-2\overline{u}_{n}u_{n+1}\\
&\quad-2\overline{u}_{n-1}u_{n}+\overline{u}_{n-1}u_{n+1}-2\overline{u}_{n+1}u_{n}+\overline{u}_{n+1}u_{n-1}.
\end{split}
\end{equation}
Before we separate terms and re-express the summation over $n$ as the sum of nine summations, we need to be sure each of the corresponding sequences is absolutely convergent. To achieve this, we first show that $\Vert u \Vert^2 = \sum_{n}|u_{n}|^2<\infty$.  This follows from the inequality:
\begin{equation}
M_\text{max}^{-1} \Vert u \Vert_M^2 \le \Vert u \Vert^2 \le M_\text{min}^{-1} \Vert u \Vert_M^2 \, ,
\label{eq:app_u_norm}
\end{equation}
from which we immediately obtain that $\Vert u \Vert < \infty$ if and only if $\Vert u \Vert_M < \infty$. Hence, $\mathcal{H}=\ell^2$ as vector subspaces of $\mathbb{C}^{\mathbb{Z}}$, and the only difference is their inner product. 

Now, let $S^{\pm}:\ell^{2}\to\ell^{2}$ denote the shift operators $(S^{\pm}u)_{n}=u_{n\pm1}$, which are bounded of norm~$1$ with respect to the standard (unweighted) $\ell^{2}$ inner product. The reason for introducing the shift operators is because a term like $\sum_{n}\overline{u}_{n}u_{n-1}$ appearing in~\eqref{eq:splitLaplacianterms} and the sum in~\eqref{eq:TuLE2n1n1} is given by the inner product $\inner{ u,S^{-}u}$, and similarly for the other terms. By the Cauchy--Schwarz inequality, we have 
\begin{equation}
\left|\inner{u,S^{\pm}u}\right|\le \lVert u\rVert \;\lVert S^{\pm}u\rVert = \lVert u\rVert^{2}<\infty.
\label{eq:app_shift}
\end{equation}
Thus, the sequences $(\overline{u_{n}}u_{n-1})$, etc., appearing in~\eqref{eq:splitLaplacianterms} are all absolutely convergent and so their sums can be calculated in any order. Therefore, 
\begin{equation}
\sum_{n}|2u_n-u_{n-1}-u_{n+1}|^2\le 16\sum_{n}|u_{n}|^2
\label{eq:app_triangle}
\end{equation}
by~\eqref{eq:splitLaplacianterms} and the triangle inequality.
Putting these arguments together, 
\begin{equation}
\lVert \hat{T}u\rVert_M^2
\le 16\left(\frac{J^2}{M_{\text{min}}}\right)\sum_{n}|u_{n}|^2
\le 16\left(\frac{J}{M_{\text{min}}}\right)^2\sum_{n} M_{n}|u_{n}|^2
=\left(\frac{4 J}{M_{\text{min}}}\right)^{2}\lVert u\rVert_M^2.
\label{eq:app_T_bounded}
\end{equation}
 
By the definition of the operator norm, this proves that $\lVert \hat{T}\rVert_M\le\frac{4J}{M_{\text{min}}}$, so that $\hat{T}$ is bounded on $\mathcal{H}$, the space of all $u\in\mathbb{C}^{\mathbb{Z}}$ satisfying $\lVert u\rVert_M<\infty$ provided that $\lim_{n\to\pm\infty}\Delta M_{n}=0$.

We next prove that $\hat{T}$ is self-adjoint with respect to the weighted inner product~\eqref{eq:inner_prod_1D}, which means $\inner{\hat{T}u,v}_M=\inner{u,\hat{T}v}_M$ for all $u,v\in\mathcal{H}$. Indeed,  
\begin{equation}
\begin{split}
\inner{ u, \hat{T}v}_M &=J\sum_{n\in\mathbb{Z}}(2\overline{u}_{n}v_{n}-\overline{u}_{n}v_{n-1}-\overline{u}_{n}v_{n+1})\\
&=J\sum_{n\in\mathbb{Z}}(2\overline{u}_{n}v_{n}-\overline{u_{n+1}}v_{n}-\overline{u_{n-1}}v_{n})\\
&=\inner{\hat{T}u,v}_M \, .
\end{split}
\label{eq:app_T_sym}
\end{equation}
The first and third equalities follow directly from the definitions of $\hat{T}$ and the weighted inner product. The second equality follows from our earlier observation that $\mathcal{H}=\ell^{2}$ as vector subspaces and the Cauchy--Schwarz inequality, which implies that the series $\sum_n\overline{u}_n v_n$, $\sum_n\overline{u}_{n+1}v_n$, and $\sum_n\overline{u}_{n-1}v_n$ are all absolutely convergent and can therefore be rearranged as shown. This proves that $\hat{T}$ is self-adjoint on $\mathcal{H}$. 

A similar rearrangement yields the identity:
\begin{equation}
\inner{ u, \hat{T}v}_M 
= J\sum_{n\in\mathbb{Z}} \overline{(u_{n+1} - u_n)} (v_{n+1} - v_n) 
= J \inner{\hat{D} u, \hat{D} v} \, ,
\label{eq:T_factor_1d}
\end{equation}
where $\hat{D}: \mathbb{C}^\mathbb{Z} \to \mathbb{C}^\mathbb{Z}$ is the difference operator $(\hat{D} u)_n = u_{n+1} - u_n$.  This is the 1d analogue of the factorization $\hat{T} =  \hat{M}^{-1} J\hat{D}^\dagger \hat{D}$ that we used in the 2d case [Eq.~\eqref{eq:difference_factorization}].

The proofs are nearly identical for the 2d $\hat{T}$ operator~\eqref{eq:T_2d}.  Eq.~\eqref{eq:app_u_norm} is identical.  The shift operators $S^{\pm}$ are extended to the analogous shift operators in~$n$ and~$m$, and then Eq.~\eqref{eq:app_shift} is identical.  The triangle inequality \eqref{eq:app_triangle} now has more terms from Eq.~\eqref{eq:T_2d}, and so one obtains a coefficient of $64$ instead of $16$, leading to an operator bound similar to Eq.~\eqref{eq:app_T_bounded} except with $\lVert \hat{T}\rVert_M\le\frac{8J}{M_{\text{min}}}$.  The self-adjointness derivation~\eqref{eq:app_T_sym} in 1d  is simply applied twice, to 1d second-difference operators along each direction.  Eq.~\eqref{eq:cond2D} is also simply applying Eq.~\eqref{eq:T_factor_1d} twice, to $\hat{D}_x$ and $\hat{D}_y$.

\section{Invariance of the essential spectrum}
\label{app:Weyl}

Our proof of localization requires that the perturbation does not change the continuous spectrum of bulk modes, which is more precisely known as the \emph{essential spectrum}~\cite{Kato1995}: the ``continuous eigenvalues'' $\lambda$ such that $\hat{T} - \lambda$ is not a Fredholm operator.  This invariance is straightforward to prove for masses satisfying our criteria that $M_n > 0$ for all $n\in\mathbb{Z}$ and $\sum_n |\Delta M_n| < \infty$ (and similarly in 2d).

Because $\hat{T}$ is self-adjoint under an inner product~\eqref{eq:inner_prod_1D} that depends on the masses, however, it is convenient to transform the problem slightly. One can express $\hat{T} =  \hat{M}^{-1} J \hat{L}$ as the product of the inverse of the (invertible) mass operator $\hat{M}$ (which multiplies elementwise by $M_n$ in 1d or by $M_{n,m}$ in 2d) and the graph Laplacian (second-difference operator) $\hat{L}$. We can then use the fact that the essential spectrum of $\hat{T}$ is equal to the essential spectrum of the \emph{operator pencil} $J\hat{L} - \lambda \hat{M}$~\cite{Nakic2016,Gernandt2020}.  (For finite matrices, this corresponds to mapping the ordinary eigenproblem $ \hat{M}^{-1} J \hat{L} u = \lambda u$ to the generalized eigenproblem $J \hat{L} u = \lambda \hat{M} u$.)   Both $\hat{M}$ and $J \hat{L}$ are self-adjoint under the \emph{unweighted} $\ell^2$ inner product $\inner{u, v} = \sum \overline{u_n} v_n$ (since $\hat{M}$ is diagonal and $J \hat{L}$ is equivalent to $\hat{T}$ with the masses set to~$1$).

The essential spectrum of an operator pencil $\hat{P}(\lambda) = J\hat{L} - \lambda \hat{M}$ is the set of $\lambda$ where $\hat{P}(\lambda)$ is not Fredholm~\cite{Nakic2016,Gernandt2020}, or equivalently where $0$ is in the essential spectrum of the operator $\hat{P}(\lambda)$.  By Weyl's theorem, the essential spectrum of $\hat{P}(\lambda)$ is unchanged if one perturbs $\hat{P}(\lambda)$ by a compact operator~\cite{Kato1995}, or equivalently if one perturbs $J\hat{L}$ and $\hat{M}$ by compact operators~\cite{Gernandt2020}.  Our mass perturbation does not change $J\hat{L}$, whereas $\hat{M}$ is changed by the operator $\Delta \hat{M}$ that multiplies elementwise by $\Delta M_n$ (or by $\Delta M_{n,m}$ in 2d).  However, $\Delta \hat{M}$ is a Hilbert--Schmidt operator and hence compact~\cite{Conway2007}: its Hilbert--Schmidt norm in the Cartesian basis is simply $\Vert \Delta \hat{M} \Vert_{\text{HS}}^2 = \sum_n |\Delta M_n|^2 \le \left( \sum_n |\Delta M_n| \right)^2 < \infty$ (and similarly in 2d).

\section{Integral bounds for sums}
\label{app:integral-bounds}

In the main text we employ the following well-known integral bounds~\cite{Wade} for a monotonically decreasing function $h(x)$:
\begin{align}
\int_{0}^{\infty}h(x) \de x \leq \sum_{n\geq0}h(n) \leq h(0)+\int_{0}^{\infty}h(x)\de x .\label{eq:1d_bounds}
\end{align}
As we prove below, a straightforward extension of this inequality to a 2d function $h(x,y)$ that is monotonically decreasing in both variables is:
\begin{align}
\int_{0}^{\infty}\int_{0}^{\infty}h(x,y)\de x \de y \leq \sum_{n,m\geq0}h(n,m) \label{eq:2d_bounds_1}
\end{align}
and 
\begin{align}
     \sum_{n,m\geq0}h(n,m) &\leq h(0,0) + \int_{0}^{\infty}h(x,0)\de x + \int_{0}^{\infty}h(0,y)\de y + \int_{0}^{\infty}\int_{0}^{\infty}h(x,y)\de x \de y.
     \label{eq:2d_bounds_2}
\end{align}
The first inequality~\eqref{eq:2d_bounds_1} is simple (left Riemann sums over-estimate integrals of decreasing functions). For the second inequality~\eqref{eq:2d_bounds_2}, define $\square_{n,m} \subset \R^2$ as the unit square with vertices $(n,m), (n+1,m), (n+1,m+1)$, and $(n,m+1)$, and let
\begin{align}
    I \coloneqq& \sum_{n,m\geq0}h(n,m) - {\int_{0}^{\infty}\int_{0}^{\infty}}h(x,y)\de x\de y \nonumber \\
    =& \Bigg[\sum_{n=0,m\geq0}h(n,m)  + \sum_{n\geq0,m=0}h(n,m) - h(0,0) +\sum_{n\geq0,m\geq0}h(n+1,m+1) \Bigg] \nonumber \\
    &\,- \sum_{n,m\geq0} \iint_{\square_{n,m}}h(x,y)\de x \de y.
\end{align}
Now use the fact that $\iint_{\square_{n,m}}dxdy = 1$ and that ${h(n+1,m+1) - h(x,y) \leq 0}$ for all $x,y\in \square_{n,m}$, yielding
\begin{align}
    I =& \sum_{n=0,m\geq0}h(n,m)  + \sum_{n\geq0,m=0}h(n,m) - h(0,0) \nonumber \\ &+\sum_{n\geq0,m\geq0} \iint_{\square_{n,m}}\left[h(n+1,m+1) - h(x,y)\right]\de x\de y \nonumber  \\
    \leq & \sum_{n=0,m\geq0}h(n,m)  + \sum_{n\geq0,m=0}h(n,m) - 
    h(0,0).
\end{align}
Finally, by rearranging and using the 1d bounds for each 1d sum, we obtain the result~(\ref{eq:2d_bounds_2}).

\end{appendices}

%
%







\bibliographystyle{vancouver}
\bibliography{biblio}

\end{document}